\documentclass[12pt,aps,prd,showpacs,notitlepage,nofootinbib]{revtex4-1}
\usepackage{amssymb,amsmath}
\usepackage{bm}
\usepackage{graphicx}
\usepackage{subfigure}
\usepackage{color}
\usepackage{amsthm}
\usepackage[update,prepend]{epstopdf}
\usepackage{hyperref}
\usepackage{natbib}
\bibpunct{[}{]}{,}{n}{,}{,}
\newcommand{\gev}{\ {\rm GeV}}
\newcommand{\tev}{\ {\rm TeV}}
\newcommand{\beq}{\begin{equation}}
\newcommand{\bea}{\begin{eqnarray}}
\newcommand{\eeq}{\end{equation}}
\newcommand{\eea}{\end{eqnarray}}

\usepackage{footnote}
\begin{document}

\title{The Relaxion: A Landscape Without Anthropics}
\author{Ann Nelson}
\email{anelson@phys.washington.edu}
\author{Chanda Prescod-Weinstein}
\email{cprescod@uw.edu}
\affiliation{{Department of Physics, University of Washington, Seattle, Washington 98195-1560}}
\date{\today}
\begin{abstract}
The relaxion mechanism provides a potentially elegant solution to the hierarchy problem without resorting to anthropic or other fine-tuning arguments. This mechanism  introduces an axion-like field, dubbed the relaxion, whose expectation value determines the electroweak hierarchy as well as the QCD strong CP violating $\bar{\theta}$ parameter. During an inflationary period,  the Higgs mass squared is selected to be negative and hierarchically small in a theory which is consistent with 't Hooft's technical naturalness criteria. However, in the original model proposed by Graham, Kaplan, and Rajendran~\cite{Graham:2015cka}, the relaxion does not solve the  strong CP problem, and in fact contributes to it, as  the coupling of the relaxion  to the Higgs field and the  introduction of a linear potential for the relaxion produces large strong CP violation. We resolve this tension by considering inflation with a Hubble scale which is above the QCD scale but below the weak scale, and estimating the Hubble temperature dependence of the axion mass. The relaxion potential is thus very different during inflation than it is today. We find that provided the inflationary Hubble scale is between the weak scale and about 3 GeV, the relaxion resolves the hierarchy, strong CP, and dark matter problems in a way that is technically natural.

\end{abstract}
\maketitle
\section{Introduction} 
Although the Standard Model represents a significant triumph for both theoretical and experimental particle physics, questions remain. One outstanding challenge is the strong CP problem, where the non-detection of an electric dipole moment for the neutron requires a tremendous fine-tuning of the strong CP violating $\bar{\theta}$ parameter~\cite{Belavin:1975fg,Jackiw:1976pf,Callan:1976je}. The most elegant solution to this problem is the Peccei-Quinn (PQ) mechanism, in which $\bar{\theta}$  is determined by the expectation value of a pseudo-Nambu-Goldstone boson known as the axion~\cite{Peccei:1977hh,Weinberg:1977ma,Wilczek:1977pj}. The  energy density  of the QCD vacuum is minimized at  the CP conserving value of $\bar{\theta}=0$. Even though the weak interactions violate $CP$, the ground state of the full theory is at  $\bar{\theta}\sim 10^{-16}$,~\cite{Ellis:1978hq} which is much smaller than the experimental limit of $\bar{\theta}<10^{-10}$  from the electric dipole moment of the neutron~\cite{Crewther:1979pi,Georgi:1986df,Dubbers:2011ns}. The coupling of the axion can be made arbitrarily weak, allowing it to escape various direct detection searches~\cite{Kim:1979if,Zhitnitsky:1980tq,Dine:1981rt}. 

Happily, for sufficiently weak coupling, the axion is inevitably produced in the early universe via the misalignment mechanism, in which case the axion can address another outstanding problem: which particle(s) constitutes the dark matter that appears to dominate cosmic structures~\cite{Abbott:1982af,Preskill:1982cy,Dine:1982ah}. Axion dark matter has become the subject of active detection searches, with the Axion Dark Matter Experiment exploring the theoretically preferred mass range~\cite{Sikivie:1983ip,Asztalos:2009yp,Rosenberg:2015kxa,Stern:2016bbw}.  

Recently, Graham, Kaplan and Rajendran (GKR) \cite{Graham:2015cka} proposed a new use for the axion: to address the electroweak hierarchy problem. While one might naively expect that the weak scale would be coincident with the Planck scale, instead Fermi's constant $G_F$, which is determined by the Higgs expectation value, is 34 orders of magnitude larger than Newton's constant $G_N$. In the Standard Model, the Higgs expectation value is determined by a mass squared parameter whose renormalized value is 34 orders of magnitude smaller than the Planck scale squared, and it is unknown why the Higgs has this mass.
 
Furthermore, the tiny value of the Higgs mass squared parameter violates the 't Hooft naturalness condition that a parameter should be very small only when a value of zero increases the symmetry of the theory~\cite{tHooft:1979rat}. The relaxion model tackles this problem by having the Higgs mass squared determined by dynamics which selects a small value. The relaxion theory does contain a small parameter, namely a tiny coupling of the relaxion to the Higgs field, but this small parameter is natural in the 't Hooft sense, as it breaks the Peccei-Quinn (PQ) symmetry.  During inflation the relaxion evolves slowly until the  Higgs mass squared parameter becomes negative. Then the Higgs develops an expectation value and the resulting back reaction stops the evolution of the relaxion and the Higgs mass squared value remains small and negative.

This relaxion mechanism satisfies 't Hooft's technically natural standard, but it also introduces new problems. In addition to potential problems with fine-tuning~\cite{DiChiara:2015euo,Fowlie:2016jlx}, the $\bar{\theta}$  angle which the axion mechanism was introduced to make small ends up being $\sim \mathcal{O}(1)$, as it is determined by equal competition between QCD dynamics, which prefers a value of zero, and  the PQ symmetry breaking coupling of the relaxion to the Higgs field. In the original axion mechanism, the minimum of the potential is $\bar{\theta} \sim 0$, but in this new relaxion picture, the potential is tilted and that is no longer the case. Thus, while the relaxion mechanism may provide an elegant resolution to the electroweak hierarchy problem, in doing so it (re)produces a new (old) problem. In fact the problem is worse, because while in the minimal standard model $\bar\theta$ is a free parameter, in the relaxion model $\bar\theta$ is dynamically determined to be large. 

GKR suggested solving this problem by having the relaxion-Higgs coupling determined by the inflaton field and having this coupling reduce dramatically post inflation, so that today the relaxion value is determined solely by QCD. However, there are no a priori technical or naturalness arguments for this particular resolution. Another possibility they suggested is to keep the QCD axion uncoupled to the Higgs, and have the relaxion be an axion-like particle for a new, nonstandard interaction, a resolution that does not share the axion's appealing necessity to resolve another challenge faced by the Standard Model.
	
In this paper, we consider the Hubble scale dependence of the relaxion potential and the resulting parameter space. The Hubble scale during inflation acts like a temperature, cutting off infrared effects, with similar effects on dynamics. Although at low temperatures (below the QCD scale) the axion mass is temperature-independent, above the QCD scale, this is not the case. We find that by relaxing the Kaplan et al. requirement that the Hubble scale remain below the QCD scale~\cite{Kobayashi:2016bue,2015JHEP...12..162B}, it is possible to find ourselves in an inflated patch of the universe where there is a high ratio between the high-temperature mass of the relaxion and the low-temperature mass. 

In this scenario, the strength of the relaxion-higgs coupling can be reduced tremendously with the relaxion mechanism still determining a hierarchically small value of the weak scale during inflation, as long as during inflation the back reaction for the Higgs vacuum expectation value of the relaxion potential has similar size to the PQ symmetry breaking scale. The effects of QCD on the relaxion potential at low temperature are then much larger after inflation than they are during inflation. Therefore, the value of the relaxion today is mostly determined by the QCD contribution to the potential, and it approximately aligns with the CP-conserving value of $\bar{\theta}$.

In Section II, we review the GKR relaxion mechanism in some detail. Section III goes on to describe how the relaxion mechanism is affected by finite temperature field theory considerations during inflation. In this section we introduce the ``landscape relaxion" in which different patches of the universe have different values of the relaxion.  We consider a statistical ensemble of inflated patches and show that a patch like ours with a small weak scale and small $\bar{\theta}$ is typical. Finally, in Section IV, we discuss our conclusions and suggest the use of the relaxion for Weinberg's anthropic landscape solution to the cosmological constant problem~\cite{Weinberg:1987dv}.

\section{Review of the (rel)axion}

The axion is a (pseudo-)scalar field $\phi$ that implements the Peccei-Quinn  (PQ) solution to the strong CP problem. The PQ mechanism addresses this Standard Model issue through a spontaneously broken global U(1) symmetry, which leads to the production of a Goldstone boson, the axion. Because the  PQ symmetry is not exact in the presence of nonperturbative QCD effects, the axion obtains a potential, which is minimized when the $\bar\theta$ parameter is zero. Having such a symmetry is technically natural as the PQ symmetry breaking is only due to nonperturbative effects which are negligible at short distances. At low temperatures, the axion potential is  of the form
\beq
V(\phi)=\Lambda^4\left(1-\cos\left(\phi/f_a\right)\right).
\eeq
 $\Lambda \sim 0.1$\,GeV is a parameter of order the QCD scale, and $f_a$ is the PQ symmetry breaking scale, often referred to as the axion decay constant. 
 
 The axion is a potential candidate for dark matter because it can be shown that the abundance of axion dark matter in the universe is determined by $f_a$ with value
\beq
\Omega_a\sim\left(f_a\over10^{11-12}\;\mbox{\gev}\right)^{7/6}.
\eeq
Uncertainty in the expression comes from the  temperature-dependence of the axion mass, as well as uncertainties in axion cosmology such as whether the PQ symmetry breaks before or after inflation, and, in the former case, on the value of the axion expectation value in our patch of the universe during inflation.

 In the relaxion scenario, the axion is repurposed to address the electroweak hierarchy problem. A PQ breaking linear term in the $\phi$ potential is introduced, as well as a coupling between $\phi$ and the Higgs field $h$. In addition, the range over which $\phi$ can vary is expanded exponentially. As the relaxion rolls down its potential, initially the Higgs mass squared is positive and the quarks are massless. With massless quarks, there is no QCD contribution to the relaxion potential. The Higgs mass squared parameter decreases until it becomes negative and the Higgs field acquires a vacuum expectation value. At this point, the quarks obtain mass and a QCD contribution to the relaxion potential turns on. The QCD contribution stops the relaxion from evolving further and the Higgs has apparently naturally arrived at the correct value. 
 
 Unlike in the original axion model, if one views the relaxion as a pseudo-Goldstone boson corresponding to spontaneous breaking of a Peccei-Quinn symmetry, the model must contain an exponentially large discrete symmetry group and the range of the field is much larger than the Planck scale~\cite{Gupta2016}. Note however that some recent work~\cite{Choi:2015fiu,Kaplan:2015fuy} shows how certain multi-field models can produce such an effective theory. The full set of relaxion couplings are
  \beq
 {\cal L}=c_1 g M^2 \phi - (M^2-g\phi) |h^2|+\left(\frac{\phi}{f_a}\right)\left(\frac{g^2}{16\pi^2}\right)G\tilde{G} \ .
 \eeq
Here $\phi$ is the relaxion, $h$ is the Higgs field, $g $ is a small coupling, $c_1$ is a positive parameter  of order one, $M$ is a high mass, and $f_a$ is similar to the usual axion  decay constant.  There is   a symmetry $\phi\rightarrow\phi+ c$ in the limit where nonperturbative QCD effects are turned off and $g\rightarrow0$.  The cutoff scale of this effective theory is taken to be of order the higgs mass parameter   $M$. The  range $\Delta\phi$ over which the relaxion can evolve is taken to be $\Delta \phi > M^2/g$, which will turn out to be much larger than $f_a$.  
The origin of the  small parameter $g$  is not addressed, but any renormalization of $g$ is proportional to $g$. Conceivably $g$ might arise from   nonperturbative breaking of the PQ symmetry from something other than QCD. As long as some high scale new physics cuts off any quadratic divergences at the scale $M$, the theory is technically natural. 

\section{Relaxion during inflation: a landscape phenomenon}

Since it is expected that during inflation perturbations in the metric can induce fluctations of the Higgs field which scale with the Hubble parameter such that per Hubble time~\cite{2005pfc..book.....M}
\beq
\delta h = \frac{H}{2\pi},
\eeq
Kaplan et al. impose a requirement on the relaxion that the classical deterministic evolution should dominate over the random thermal wandering  in a Hubble time, 
\beq \label{lowH}
H<\Lambda_{QCD}, \  H < (g M^2)^{1/3}.
\eeq
Then using  
\beq
g M^2 f_a\sim m_a^2 f_a^2
\eeq and 
\beq \label{Hdom}
H>M^2/M_{Pl}
\eeq (so that the inflationary energy density was greater than the change in the energy density due to $\phi $ rolling) they concluded
\beq
M<\Big(\frac{m_a^2 f_a^2 M_{Pl}^3}{f_a}\Big)^{1/6}\sim 10^7 \Big(\frac{10^9 \gev}{f_a}\Big)^{1/6}.
\eeq
With this constraint, the Hubble scale during inflation is necessarily below the QCD scale. Phenomenologically this is consistent with current constraints on the tensor-to-scalar ratio from data~\cite{2016AA...594A..13P}. However in this scenario, the  $\bar\theta$  parameter is of order one today, in contradiction with laboratory experiments~\cite{PhysRevD.92.092003}. 

\subsection{Addressing the CP Problem}

\begin{figure}[ht]
\centering
\subfigure[GKR relaxion during inflation]{%
\includegraphics[width=.45\textwidth]{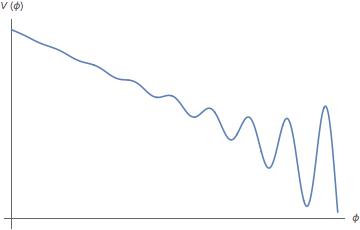}
\label{fig:subfigure1}}
\quad
\subfigure[landscape relaxion during inflation]{%
\includegraphics[width=.45\textwidth]{badtheta.jpg} 
\label{fig:subfigure2}}
\subfigure[late universe GKR relaxion]{%
\includegraphics[width=.45\textwidth]{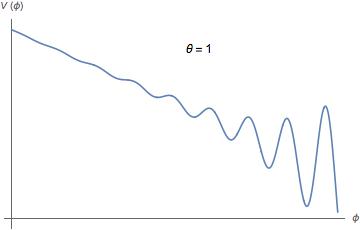}
\label{fig:subfigure3}}
\subfigure[late universe landscape relaxion]{%
\includegraphics[width=.45\textwidth]{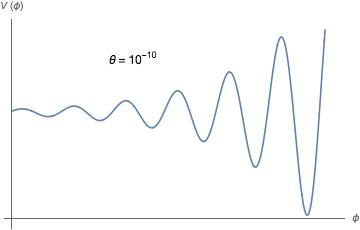} 
\label{fig:subfigure4}}
\caption{\small \baselineskip 11pt
This figure is primarily to give readers an intuition for the similarities and differences in the two models. In the landscape case the QCD contribution to the potential is greatly reduced during inflation, and the explicit PQ symmetry breaking is also much smaller. The scales on the left and right differ drastically. In both cases, during inflation,  the relaxion dynamics are affected by both the PQ breaking parameter and QCD nonperturbative affects. For the GKR case the relaxion potential during inflation is almost the same as it is today.  For the landscape case,  the QCD contribution to the potential  is vastly larger  today  than it was during inflation, so the scale used for depicting $V(\phi)$ is increased accordingly for the late universe. Top left: during inflation, $H \lesssim \Lambda_{QCD}$, $\theta \sim1$; Bottom left: late universe, $H \sim 0$, $\theta \sim 1$; Top right: during inflation, $M_{W} > H >> \Lambda_{QCD}$, $\theta \sim 1$; Bottom right: late universe, $H  \sim 0$, $\theta \sim 0$.   }
\label{fig:figure}
\end{figure}

To address the relaxion's strong CP problem, we first note that the effects of the horizon during inflation has similar effects on the dynamics as does a finite temperature (see, e.g. ref.~\cite{Spradlin:2001pw} for a review). Therefore, we estimate the effects of a high Hubble scale by using the finite temperature computation of the relaxion mass evaluated at a temperature of $H$. We then confront the relaxion's CP problem by relaxing the requirement of eq.~\ref{lowH} and do not try to suppress the landscape of final relaxion values. In the process, we do not invoke any anthropic principle for the weak scale, but rather we examine the parameters for which the majority of vacua agree with observation in that they have a hierarchically small weak scale. We call the result the ``landscape relaxion.''
 
 When the Higgs field $h$ has a positive mass squared, it does not have a vacuum expectation value (VEV), and the quarks are massless. Massless quarks greatly suppress the effects of QCD instantons, which give the relaxion its mass. Neglecting the variation in the Hubble scale during inflation, and including the effects of QCD instantons, we take the relaxion potential to be
\beq
V(\phi) = -g M^2 \phi + (M^2-g\phi) |h^2| - \frac{f(v)}{b}(m_a^2 f_a^2)\cos\left(\frac{\phi}{f_a}\right)
 \eeq
 where the zero temperature value of the relaxion mass is $m_a$. The factor $b$ is the ratio of the zero temperature value of the relaxion mass squared to the value of the mass squared during inflation. We assume that the non-zero temperature value of the relaxion mass is given per~\cite{Preskill:1982cy} and~\cite{Gross:1980br}.   We parameterize the back reaction of the Higgs VEV on the relaxion potential by the function $f(v)$ which is a function of the Higgs VEV $v$, noting that $v$ is a function of $\phi$.  We take $f(v)=1$ when the Higgs VEV takes its final value.  When  the Higgs mass squared is positive, which happens when $ (M^2-g\phi) >0$, we neglect the tiny correction to the relaxion potential and take $f(v)$ to be zero.
 
 We give a qualitative description of the relaxion dynamics as follows. At the start of inflation we have $(M^2-g\phi)>0$,  a positive Higgs mass squared, and $v=0$. Every Hubble time, $H^{-1}$,  $\phi$ wanders randomly by an amount of order $H/(2\pi)$. In addition   the expectation value of $\phi$ evolves classically. When $v=0$, the expectation value of $\phi$ is    pushed by the $-g M^2 \phi $ term in the potential and changes  by an amount $g M^2/H^2$ per Hubble time.  After $N\sim H^2/g^2$ Hubble times, the relaxion average value has  changed by  $\sim (M^2/g)$, as needed for the average value of the Higgs mass squared to be negative. Using a random walk model of $\Delta\phi\sim H$ per Hubble time gives a spread in the value of $\phi $ of order $\sqrt{N}H\sim H^2/g$. Thus after $H^2/g^2$ Hubble times the Higgs mass squared has evolved to $\sim 0\pm H^2$.  We assume $H$ to be much smaller than the value of the Higgs mass in our patch of the universe. After $\sim (1+ H^2/M^2)(H^2/g^2) $ e-folds, most of the relaxion values are such that the Higgs mass squared is slightly negative. 
 
 For classical evolution, as was assumed by GKR, the slow roll of the average expectation value stops due to back reaction when $M^2 g= f(v) (m_a^2 f^2_a)/(f_a b)$. We take this to happen by definition at $(M^2-g\phi)=-m_h^2$, where $-m_h^2$ is the value of the Higgs mass parameter in our world, which happens after about $\sim (1+ m_h^2/M^2)(H^2/g^2)$ e-folds. If the back reaction happens in a similar manner when $H$ is large, the universe consists of a tremendous number of causally disconnected patches, each with a different value of $\phi$. However, due to the small value of $g$, at the time when the back reaction takes place, the spread in the value of the Higgs mass squared parameter is small, of order $H^2$.  The probability distribution will continue to spread  until the end of inflation, with the variance in the weak scale of order $g\sqrt{H^3 t}$. Therefore 
as long as inflation does not last for more than $m_h^4/(g^2 H^2) $ e-folds, the variance in the weak scale is less than the weak scale.

We now examine this picture more quantitatively.
 
 \subsection{The Probability Distribution of Relaxion Values}\label{probability}
The large spread in relaxion values is not in accord with a deterministic classical picture of the dynamics. We may examine the back reaction using the Fokker-Planck equation,  as described in the context of Higgs dynamics in ref.~\cite{Espinosa:2007qp}. 
 \beq\label{fp}
 \frac{\partial P}{\partial t}=
 \frac{\partial}{\partial \phi} \left( \frac{H^3}{8 \pi^2} \frac{\partial P}{\partial \phi} + \frac{V' P}{3 H} \right)
 \eeq
  Here   $P(\phi,t)$ is the probability of finding value $\phi$ for the relaxion at time $t$.

 For  constant $H$ and constant $V'$, just as indicated by our qualitative discussion, a solution to eq.~\ref{fp}  is a spreading   gaussian, with the width growing as $t^{1/2}$, and mean value slowly rolling down the potential. We define   $P_0(\phi,t)$ to be the probability of finding value $\phi$ for the relaxion at time $t$, for the case where the initial distribution is a delta function and the back reaction from QCD is turned off, so that $V'=-g M^2$.
 \beq
 \label{fpsol}
P_0(\phi,t)=\frac{
                           \sqrt{2\pi }
                           }{
                              \sqrt{H^3t}
                              }
                              e^{
                              -2\pi^2
                              \frac{
                              \left( \phi - g M^2 t/(3 H)\right)^2
                              }{
                              H^3 t}
                              }
 \eeq

   GKR assumed for an initial distribution with a small spread in the values of $\phi$, $P$ stops evolving when most of the values of $\phi$ are in a regime where $V'\sim 0$.  Because $V'$ is an oscillating function of $\phi$, this approximation requires that $P$   can be approximated by a delta function of $\phi$.   
   
  A qualitative picture of the dynamics when the center of the distribution $P_0$ reaches the regime where the QCD contribution to the potential   is important is as follows.  $P(\phi,t)$ will evolve to become larger in regions where the potential is locally minimized with respect to $\phi$ and smaller in regions  of local maxima.  Due to this back reaction,  the expectation value of $\phi$ will stop increasing. 

In order to give a more quantitative treatment, as we do not know how to find an exact solution of the Fokker-Planck equation in the presence of the QCD term,  we treat  the QCD contribution as a perturbation.  
We take  $V'$  to be  $-g M^2 + \lambda \epsilon'(\phi)$, where $\epsilon'=- \frac{f(v)}{ f_a b}(m_a^2 f_a^2)\sin(\frac{\phi}{f_a})$. We take   $P=P_0+\lambda p(\phi,t)$ and treat $\lambda$ as an expansion parameter. Collecting terms which are linear in $\lambda$ we find $p$ satisfies

\bea
\frac{\partial p}{\partial t}&=&\frac{\partial}{\partial \phi} \left( \frac{H^3}{8 \pi^2} \frac{\partial p}{\partial \phi} + 
\frac{\epsilon'(\phi) P_0(\phi,t)}{3 H} \right) \eea

We see that equation for the perturbation has the same form as the heat equation with a driving term, also known as the forced heat equation, and we may solve it using a Green's function technique, which results in an explicit albeit complicated integral.   
 The appropriate Greens function $G$ is the solution to the homogenous equation multiplied by a step function:
\beq 
G(t,t_0,\phi,\phi_0)= \frac{ \sqrt{2\pi } \theta(t-t_0)}{\sqrt{H^3(t-t_0)} }e^{ -2\pi^2 \frac{ (\phi-\phi_0  )^2}{H^3 (t-t_0)} }
\eeq 
so we take the integral of this Green's function multiplied by the driving term

\bea p(\phi,t)&=&\int_0^t dt_0 \int^\infty_{-\infty} d\phi_0\frac{\sqrt{2\pi }}{\sqrt{H^3(t-t_0)}}e^{ -2\pi^2 \frac{ (\phi-\phi_0  )^2}{H^3 (t-t_0)} } \frac{\partial}{\partial\phi_0}  \left(\frac{\epsilon'(\phi_0) P_0(\phi_0,t_0)}{3 H}\right)\cr
&=&\int_0^t dt_0 \int^\infty_{-\infty} d\phi_0\frac{- (2\pi)^{\frac52}(\phi-\phi_0)}{ H^{\frac92}(t-t_0)^{3/2}  }e^{ -2\pi^2 \frac{ (\phi-\phi_0  )^2}{H^3 (t-t_0)} }  \frac{\epsilon'(\phi_0) P_0(\phi_0,t_0)}{3 H}  \cr
&=&
\int_0^t dt_0 \int^\infty_{-\infty} d\phi_0
\frac{-8\pi^3(\phi-\phi_0)\epsilon'(\phi_0) }{ 3 H^7(t-t_0)^{\frac32}t_0^{\frac12} }
e^{ -2\pi^2 \left(
\frac{ (\phi-\phi_0  )^2}{H^3 (t-t_0)}+\frac{( \phi_0 - g M^2 t_0/(3 H) )^2}{H^3t_0} \right)
}  \cr&\approx&\int_0^t dt_0 \int^\infty_{-\infty} d\phi_0
\frac{-8\pi^3(\phi-\phi_0)\epsilon'(\phi_0) }{ 3 H^7(t-t_0)^{\frac32}t_0^{\frac12} }
e^{ -2\pi^2 \left(
\frac{ (\phi_0-\phi t_0/t)^2 t+ \phi^2(t-t_0)t_0/t  }{H^3 (t-t_0)t_0}  \right)}\cr
&=&
\int_0^t dt_0 \int^\infty_{-\infty} d\phi_0
\frac{-8\pi^3(\phi(1+t_0/t)-\phi_0)\epsilon'(\phi_0+\phi t_0/t) }{ 3 H^7(t-t_0)^{\frac32}t_0^{\frac12} }
e^{ -2\pi^2 \left(\frac{ (\phi_0)^2 t+ \phi^2(t-t_0)t_0/t  }{H^3 (t-t_0)t_0}  \right)}
  \eea
where we have changed variables $\phi_0 \rightarrow \phi_0 + \phi t_0/t$ and dropped terms which are proportional to $g$ in the fourth line since we are mainly interested in studying the effects of the back reaction from QCD. Note that this means we are taking the distribution function to be localized at $\phi=0$ at $t=0$, an acceptable assumption since we may always shift $t$ and $\phi$ to an arbitrary value. It is also acceptable to ignore the evolution of the mean value of $P_0$ during the back reaction    because as we show below, the back reaction sets in quickly enough that we may ignore the terms proportional to $g$ during this process. 

This equation is exactly solvable, but the full solution is not needed to serve our purposes. Rather than sharing the exact solution, we focus on extracting information related to whether the time scale for the back reaction to become significant is short enough, such that the approximation of neglecting the motion of the center of the probability distribution is sufficiently accurate.
The term $\epsilon'$ oscillates  and the concern is that the integrand will be greatly suppressed by cancellations. Taking $\epsilon'= q \sin(\phi/f_a)$, with $q=m_a^2 f_a/b$, we obtain
\beq
p(\phi,t) \approx \int_0^t dt_0 \int^\infty_{-\infty} d\phi_0
\frac{-8\pi^3 q(\phi(1+t_0/t)-\phi_0)\sin\left(\frac{\phi_0+\phi t_0/t}{f_a}\right) }{ 3 H^7(t-t_0)^{\frac32}t_0^{\frac12} }
e^{ -2\pi^2 \left(\frac{ (\phi_0)^2 t+ \phi^2(t-t_0)t_0/t  }{H^3 (t-t_0)t_0}  \right)}
\eeq

\beq
= \int_0^t dt_0 \frac{q \sqrt{2\pi}e^{\left(\frac{H^3 t_0 \left(t_0-t\right)}{8 \pi ^2 f_a^2}-\frac{2 \pi ^2 \phi^2}{H^3 t}\right)} \left(H^3  \left(t-t_0\right) t_0 \cos \left(\frac{\phi t_0}{tf_a}\right)-4 \pi ^2 f_a \phi \left(t+t_0\right) \sin \left(\frac{\phi t_0}{tf_a}\right)\right)}{3 t^{\frac32}(t-t_0)f_a H^{\frac{11}{2}} }                                                                                           \eeq

  We assume the back reaction stops the expectation value of $\phi$ from evolving further once $p$ becomes of similar magnitude to $P_0$. We wish to examine whether the time scale for this to happen is short or long compared with the very long timescale $t_{\rm higgs}$ over which the Higgs VEV changes by order one, which is $t_{\rm higgs}\sim    H m_h^2/(g^2 M^2)$.    We examine the value of the integral at $\phi=0$ since this is a relatively arbitrary value  as well as being a point at which the cosine is extremized and the distribution is localizing. We have checked numerically that our result does not significantly depend   on $\phi$. At $\phi=0$ we get a relatively simple expression:
\beq \label{expansionphi}
   p(0,t) \approx  \frac{2\pi^2  q e^{-\frac{H^3 t}{32\pi^2 f_a^2}}{\rm  erfi}\left(\sqrt{\frac{H^3 t}{32\pi^2 f_a^2}}\right) }{ 3  H^4 }\eeq
   while 
   \beq P_0(0,t)\sim \frac{1}{\sqrt{H^3 t}} \ .\eeq
   For small $t$, we Taylor expand eq.~\ref{expansionphi}
   \beq\frac{  p(0,t) }{P_0(0,t)}\sim \frac{q t  }{ H f_a } \eeq
   where we have dropped $\mathcal{O}(1)$ factors for simplicity.
  The ratio  grows approximately linearly in $t$ and becomes of $\mathcal{O}(1)$ at a time $t_{\rm back\ reaction}$
  \beq
 t_{\rm back\ reaction} \sim \frac{f_a H}{q } \ . \eeq 
   
  We compare the (long) back reaction time scale with the (very long) time scale $ t_{\rm higgs}$   and obtain
\beq\frac{ t_{\rm back\ reaction} }{ t_{\rm higgs} }\sim \frac{f_a g}{ m_h^2}\ . \eeq   The numbers we will arrive at in the next section give a small value for $g$ relative to all relevant scales, and the timescale ratio is generally less than $\sim 10^{-26}$.

\subsection{ A technically natural solution to the Strong CP problem}
At the end of inflation the Hubble parameter decreases. As the universe subsequently cools,   the QCD contribution to the relaxion potential increases.  The parameter $b$, which is defined to be the ratio of the QCD contribution to be relaxion potential today to the QCD contribution during inflation,  determines the value of $\bar\theta$ as follows. During inflation the QCD contribution is comparable to the explicit symmetry breaking. Once the universe has cooled, the explicit symmetry breaking remains the same, but the QCD contribution is larger by a factor of $b$. Since the QCD contribution is minimized at $\bar\theta=0$,  provided $b>10^{10}$, the strong CP problem is solved (see Figure 1). Such a value can occur naturally if, using finite temperature to estimate the QCD contribution to the relaxion potential during inflation, the Hubble scale during inflation lies in the approximate range 
 \beq \label{Hrange} 3 \gev < H<100 \gev.
\eeq  
The lower bound is determined by estimating the temperature at which the  QCD contribution to the potential is at least $10^{10}$ times smaller than the zero temperature value, and the upper bound is required so that the Hubble temperature does not prevent the Higgs field from gaining an expectation value during inflation. Such low-scale inflationary models appear to be required by the relaxion but have also been considered relevant in other cosmological contexts~\cite{Moroi:1993mb,Kawasaki:1994af,Bolz:2000fu,German:2001tz,Choi:2016luu,Evans:2017bjs}. 
  
  The tiny size of $g$ and enormous range of $\phi$ required to make this scenario work may seem rather extravagant. The value of $g$ is given by
\beq
g\sim \frac{m_a^2 f^2_a}{ M^2 f_a b}\sim 10^{-30} {\rm MeV} \frac{(10 \tev)^2}{M^2}\frac{10^{10}}{b}.
\eeq
(Note that at zero temperature $f_a^2 m_a^2 \sim (80$ MeV$)^4$ for the QCD axion independent of $m_a$.) This extremely tiny number is, however natural, in the sense that it violates the Peccei-Quinn symmetry which is otherwise only violated by UV insensitive nonperturbative QCD effects. Thus radiative quantum corrections to $g$ are proportional to $g$.  With $H> 3 \gev$, the number of required e-foldings during inflation is a large number, at least $\sim 10^{33}$. The range of $\phi$ is also enormous. With $M\sim 10 \tev$, $\phi$ must change by $\sim 10^{44}$ GeV. An upper limit on M comes from combining equations \ref{Hdom} and \ref{Hrange}, which gives an upper limit of $M < 10^{11} \gev$.

In addition, $\phi$ is spread over about at least $10^{20}$ distinct vacua and perhaps many more. These are all similar, as the weak scale only varies by a fraction $\sim H/m_h$  over all of  these, and the strong CP parameter is of order $1/b$ in all of them. For larger $H$, $g$ is tinier and $b$ larger. We note that for $H$ just below the weak scale, the Higgs expectation value during inflation will be much smaller than it is today, and all 6 quark masses play a role in suppressing instantons, so extremely large values of $b$ are possible in this limit. Such extreme numbers are the price to pay to avoid  introducing   arbitrary couplings or new particles to resolve the strong CP problem with the relaxion.

\subsection{Taking Measure of Our Inflationary Universe}
In~\cite{Graham:2015cka}, the authors state that the scale of inflation must observe the constraint $H < (g M^2)^{1/3}$ in order to avoid needing to address the measure problem, which naturally arises in the context of eternal inflation~\cite{LINDE1986395}, the attractor configuration for most inflationary pictures. In the eternal inflation scenario, an infinite number of inflationary regions are continuously produced, and it is difficult to make statements about probabilities such as the one estimated in~\ref{probability}. Specifically in the case of the landscape relaxion, our Fokker-Planck treatment neglects the effects the relaxion energy density on the expansion rate after inflation has ended.  Regions with higher energy density will expand faster and therefore make up an increasingly large percentage -- insofar as one can be calculated -- of the physical spacetime. Thus, the question arises of whether a similar probability estimate to the one above can be made to address whether we are likely to end up in the region with electroweak symmetry breaking.

For these reasons, the original attempt to avoid the measure problem is understandable, although it introduces different challenges. This includes requires an unusually low Hubble scale that is many orders of magnitude below what is usually modeled in typical inflationary theories, and the requirement that the linear term in the relaxion potential turn off after inflation. However, as noted above, low-scale inflation can be workable, so this is not a catastrophic change to early universe dynamics. Yet another problem of the original relaxion model is that which we address in this paper, resolving the Strong CP and dark matter problems. The initially proposed solution involves introducing a coupling to the inflaton which is only motivated by the need to address the value of $\theta$ rather than any fundamental symmetry considerations. By introducing a solution that raises the minimum Hubble scale in order to take the temperature-dependence of the axion mass into account, we reintroduce a landscape which suggests that we should also address the measure problem.

While it may seem like we are now in an impossible situation, where neither the GKR relaxion nor the landscape relaxion provide satisfactory solutions to both the electroweak hierarchy problem and the strong CP problem without introducing unattractive cosmological features. Yet, we contend that there are promising solutions to the measure problem which can potentially offer a way out (e.g.~\cite{PhysRevD.50.730,PhysRevD.52.3365,PhysRevD.60.023501}) and which have distinct observational signatures~\cite{PhysRevD.78.063520}. While we leave a more complete picture to future work, we note here that the measure problem can be resolved by one particularly promising mechanism known as the scale-factor cutoff measure~\cite{PhysRevD.78.063520}.

In the scale-factor cutoff scenario, the relative probability of any two events A and B occurring is calculated in the following manner:
\beq
{p(A) \over p(B)} \equiv \lim_{t_c \to \infty} {{n\bigl(A,
\Gamma(\Sigma, t_c)\bigr)} \over {n\bigl(B, \Gamma(\Sigma,
t_c)\bigr)}} \ ,
\eeq
where $n(A,\Gamma)$ and $n(B,\Gamma)$ are the number of events of
types $A$ and $B$, respectively, in a spacetime region we call $\Gamma$, which is constructed from a hypersurface $\Sigma$ and a time coordinate $t$. The time coordinate has been implicitly introduced via the cutoff $t_{c}$, which selects a finite spacetime region before the limit of $t_{c}$ is taken to infinity. Thus we start with an inflating spatial region $\Sigma$ and follow its evolution along geodesics orthogonal to it. Following the global time cutoff mechanism~\cite{PhysRevD.50.730,PhysRevD.51.429,PhysRevD.52.6730}, the cutoff time $t_{c}$ is introduced, but ultimately taken to infinity. Therefore we calculate probabilities by averaging over the spacetime volume that exists in a comoving region measured in time $t$, which ultimately goes to infinity. It can be shown that these probabilities are independent of which hypersurface $\Sigma$ is chosen.

 The scale-factor cutoff measure is our preferred mechanism for addressing the problem because it provides a method for stating a numerical probability given a correct theory of quantum gravity while not suffering from various difficult to resolve issues that arise in other mechanisms, for example the problem of bias toward young observers, which introduces another problem known as ``the youngness paradox.'' In addition, it has been shown in~\cite{PhysRevD.78.063520} that this mechanism provides a compelling resolution to the cosmological constant problem, so it is reasonable to expect that the same will be true for the landscape relaxion. We leave to future work a detailed calculation which shows this.

\section{In Conclusion, A Landscape}
In the original relaxion paper, GKR imposed the restriction $H<\Lambda_{QCD}$, giving as an explanation that they wished to avoid a landscape of possible values for the relaxion field. Unfortunately that prediction gives the wrong answer for the QCD strong CP parameter $\bar\theta$. GKR proposed some solutions, e.g. having the PQ breaking parameter become exponentially small when the inflaton turns off, but do not propose a symmetry-protected reason for how this could happen. They also considered having the relaxion be distinct from the QCD axion. 

In the current work, we conclude that if we relax the upper bound on $H$ given in the original relaxion paper, at the price of introducing a small but symmetry-protected number and assuming an even longer period for inflation with an even larger range for the relaxion, then the relaxion may be used to also solve the strong CP problem and provide dark matter.  When the inflationary Hubble scale is higher than $\Lambda_{QCD}$ (but still below the weak scale), then the Hubble scale acts like a temperature in suppressing the effects of large QCD instantons. Small instantons means that the  explicit PQ breaking is also much smaller while maintaining sufficient back reaction to implement the relaxion mechanism. 

Once the post-inflationary universe has cooled to well below the QCD scale, the instanton effects become much larger, by a factor $b$, and dominate the relaxion potential. The 
 zero temperature value of $\bar\theta$ comes out to order $1/b$, so that the relaxion solves the strong CP problem provided that $b> 10^{10}$. The restriction that $b>10^{10}$ places a lower bound on $H$ of order 3 GeV, which is above the upper bound of GKR. In this higher $H$ scenario, the expectation value of the relaxion is spread over an exponentially large number of local minima of  the relaxion potential.   However the spread in the value of the weak scale is still   small.  Thus the relaxion provides a natural mechanism for the production of a landscape of universes with similar values of the weak scale but different vacuum energies. 

While our relaxion does populate a landscape of vacua, we have not invoked any anthropic arguments for the strong CP, weak hierarchy, or dark matter problems. However, we still have a finely tuned cosmological constant in most or all of these vacua.  We conclude by succumbing to the temptation to remark that the relaxion landscape could allow a way to address the cosmological constant problem via Weinberg's  anthropic landscape~\cite{Weinberg:1987dv}.  The argument of Weinberg is that only those vacua with energy small enough to allow for structure formation before the expansion of the universe accelerates will have galaxies, stars and observers. The change in the value of the energy density between adjacent metastable vacua is $M^2 g f_a\sim m_a^2 f_a^2 /b$.  With an  extreme value  of $b$, the energy differences between vacua with similar particle properties  are smaller than the size of the cosmological constant.  Such a large value of $b$ could be possible in the case where the inflationary Hubble scale is not too far below the weak scale. This would at least reduce the scope of the cosmological constant problem to ensuring that the range of energies scanned by  metastable vacua includes the value zero.

\bigskip
\begin{center}{\bf Acknowledgments}\end{center}


The authors would like to thank our anonymous referees, Sarah Ballard, Jolyon Bloomfield, Ki-woon Choi, Alan Guth, Mark Hertzberg, David Kaiser, David B. Kaplan, Robert McNees, Surjeet Rajendran, Leo Stein, Scott Trager, Sarah Tuttle, and Anna Watts for helpful discussions. We thank the Aspen Center for Physics administrative and facilities staff who are supported by National Science Foundation grant PHY-1066293 for generously providing work space where some of this work was performed; the Department of Energy for partial support under grant number DE-SC0011637; and the University of Washington College of Arts \& Sciences and all its administrative and facilities staff. 

\bibliography{relaxion_landscape_5.0.bbl}

\providecommand{\href}[2]{#2}\begingroup\raggedright\begin{thebibliography}{10}

\bibitem{Graham:2015cka}
P.~W. Graham, D.~E. Kaplan and S.~Rajendran, \emph{{Cosmological Relaxation of
  the Electroweak Scale}},
  \href{https://doi.org/10.1103/PhysRevLett.115.221801}{\emph{Phys. Rev. Lett.}
  {\bfseries 115} (2015) 221801},
  [\href{https://arxiv.org/abs/1504.07551}{{\ttfamily 1504.07551}}].

\bibitem{Belavin:1975fg}
A.~A. Belavin, A.~M. Polyakov, A.~S. Schwartz and {\relax Yu}.~S. Tyupkin,
  \emph{{Pseudoparticle Solutions of the Yang-Mills Equations}},
  \href{https://doi.org/10.1016/0370-2693(75)90163-X}{\emph{Phys. Lett.}
  {\bfseries B59} (1975) 85--87}.

\bibitem{Jackiw:1976pf}
R.~Jackiw and C.~Rebbi, \emph{{Vacuum Periodicity in a Yang-Mills Quantum
  Theory}}, \href{https://doi.org/10.1103/PhysRevLett.37.172}{\emph{Phys. Rev.
  Lett.} {\bfseries 37} (1976) 172--175}.

\bibitem{Callan:1976je}
C.~G. Callan, Jr., R.~F. Dashen and D.~J. Gross, \emph{{The Structure of the
  Gauge Theory Vacuum}},
  \href{https://doi.org/10.1016/0370-2693(76)90277-X}{\emph{Phys. Lett.}
  {\bfseries B63} (1976) 334--340}.

\bibitem{Peccei:1977hh}
R.~D. Peccei and H.~R. Quinn, \emph{{CP Conservation in the Presence of
  Instantons}}, \href{https://doi.org/10.1103/PhysRevLett.38.1440}{\emph{Phys.
  Rev. Lett.} {\bfseries 38} (1977) 1440--1443}.

\bibitem{Weinberg:1977ma}
S.~Weinberg, \emph{{A New Light Boson?}},
  \href{https://doi.org/10.1103/PhysRevLett.40.223}{\emph{Phys. Rev. Lett.}
  {\bfseries 40} (1978) 223--226}.

\bibitem{Wilczek:1977pj}
F.~Wilczek, \emph{{Problem of Strong p and t Invariance in the Presence of
  Instantons}}, \href{https://doi.org/10.1103/PhysRevLett.40.279}{\emph{Phys.
  Rev. Lett.} {\bfseries 40} (1978) 279--282}.

\bibitem{Ellis:1978hq}
J.~R. Ellis and M.~K. Gaillard, \emph{{Strong and Weak CP Violation}},
  \href{https://doi.org/10.1016/0550-3213(79)90297-9}{\emph{Nucl. Phys.}
  {\bfseries B150} (1979) 141--162}.

\bibitem{Crewther:1979pi}
R.~J. Crewther, P.~Di~Vecchia, G.~Veneziano and E.~Witten, \emph{{Chiral
  Estimate of the Electric Dipole Moment of the Neutron in Quantum
  Chromodynamics}}, \href{https://doi.org/10.1016/0370-2693(80)91025-4,
  10.1016/0370-2693(79)90128-X}{\emph{Phys. Lett.} {\bfseries 88B} (1979) 123}.

\bibitem{Georgi:1986df}
H.~Georgi, D.~B. Kaplan and L.~Randall, \emph{{Manifesting the Invisible Axion
  at Low-energies}},
  \href{https://doi.org/10.1016/0370-2693(86)90688-X}{\emph{Phys. Lett.}
  {\bfseries B169} (1986) 73--78}.

\bibitem{Dubbers:2011ns}
D.~Dubbers and M.~G. Schmidt, \emph{{The Neutron and Its Role in Cosmology and
  Particle Physics}},
  \href{https://doi.org/10.1103/RevModPhys.83.1111}{\emph{Rev. Mod. Phys.}
  {\bfseries 83} (2011) 1111--1171},
  [\href{https://arxiv.org/abs/1105.3694}{{\ttfamily 1105.3694}}].

\bibitem{Kim:1979if}
J.~E. Kim, \emph{{Weak Interaction Singlet and Strong CP Invariance}},
  \href{https://doi.org/10.1103/PhysRevLett.43.103}{\emph{Phys. Rev. Lett.}
  {\bfseries 43} (1979) 103}.

\bibitem{Zhitnitsky:1980tq}
A.~R. Zhitnitsky, \emph{{On Possible Suppression of the Axion Hadron
  Interactions. (In Russian)}}, {\emph{Sov. J. Nucl. Phys.} {\bfseries 31}
  (1980) 260}.

\bibitem{Dine:1981rt}
M.~Dine, W.~Fischler and M.~Srednicki, \emph{{A Simple Solution to the Strong
  CP Problem with a Harmless Axion}},
  \href{https://doi.org/10.1016/0370-2693(81)90590-6}{\emph{Phys. Lett.}
  {\bfseries B104} (1981) 199--202}.

\bibitem{Abbott:1982af}
L.~F. Abbott and P.~Sikivie, \emph{{A Cosmological Bound on the Invisible
  Axion}}, \href{https://doi.org/10.1016/0370-2693(83)90638-X}{\emph{Phys.
  Lett.} {\bfseries B120} (1983) 133--136}.

\bibitem{Preskill:1982cy}
J.~Preskill, M.~B. Wise and F.~Wilczek, \emph{{Cosmology of the Invisible
  Axion}}, \href{https://doi.org/10.1016/0370-2693(83)90637-8}{\emph{Phys.
  Lett.} {\bfseries B120} (1983) 127--132}.

\bibitem{Dine:1982ah}
M.~Dine and W.~Fischler, \emph{{The Not So Harmless Axion}},
  \href{https://doi.org/10.1016/0370-2693(83)90639-1}{\emph{Phys. Lett.}
  {\bfseries B120} (1983) 137--141}.

\bibitem{Sikivie:1983ip}
P.~Sikivie, \emph{{Experimental Tests of the Invisible Axion}},
  \href{https://doi.org/10.1103/PhysRevLett.51.1415}{\emph{Phys. Rev. Lett.}
  {\bfseries 51} (1983) 1415--1417}.

\bibitem{Asztalos:2009yp}
{\scshape ADMX} collaboration, S.~J. Asztalos et~al., \emph{{A SQUID-based
  microwave cavity search for dark-matter axions}},
  \href{https://doi.org/10.1103/PhysRevLett.104.041301}{\emph{Phys. Rev. Lett.}
  {\bfseries 104} (2010) 041301},
  [\href{https://arxiv.org/abs/0910.5914}{{\ttfamily 0910.5914}}].

\bibitem{Rosenberg:2015kxa}
L.~J. Rosenberg, \emph{{Dark-matter QCD-axion searches}},  in \emph{{Sackler
  Colloquium: Dark Matter Universe: On the Threshhold of Discovery Irvine, USA,
  October 18-20, 2012}}, 2015,
  \href{https://doi.org/10.1073/pnas.1308788112}{DOI}.

\bibitem{Stern:2016bbw}
I.~Stern, \emph{{ADMX Status}},  in \emph{{Proceedings, 38th International
  Conference on High Energy Physics (ICHEP 2016): Chicago, IL, USA, August
  3-10, 2016}}, 2016, \href{https://arxiv.org/abs/1612.08296}{{\ttfamily
  1612.08296}},
  \href{http://inspirehep.net/record/1506408/files/arXiv:1612.08296.pdf}{http://inspirehep.net/record/1506408/files/arXiv:1612.08296.pdf}.

\bibitem{tHooft:1979rat}
G.~'t~Hooft, \emph{{Naturalness, chiral symmetry, and spontaneous chiral
  symmetry breaking}},
  \href{https://doi.org/10.1007/978-1-4684-7571-5_9}{\emph{NATO Sci. Ser. B}
  {\bfseries 59} (1980) 135--157}.

\bibitem{DiChiara:2015euo}
S.~Di~Chiara, K.~Kannike, L.~Marzola, A.~Racioppi, M.~Raidal and C.~Spethmann,
  \emph{{Relaxion Cosmology and the Price of Fine-Tuning}},
  \href{https://doi.org/10.1103/PhysRevD.93.103527}{\emph{Phys. Rev.}
  {\bfseries D93} (2016) 103527},
  [\href{https://arxiv.org/abs/1511.02858}{{\ttfamily 1511.02858}}].

\bibitem{Fowlie:2016jlx}
A.~Fowlie, C.~Balazs, G.~White, L.~Marzola and M.~Raidal, \emph{{Naturalness of
  the relaxion mechanism}},
  \href{https://doi.org/10.1007/JHEP08(2016)100}{\emph{JHEP} {\bfseries 08}
  (2016) 100}, [\href{https://arxiv.org/abs/1602.03889}{{\ttfamily
  1602.03889}}].

\bibitem{Kobayashi:2016bue}
T.~Kobayashi, O.~Seto, T.~Shimomura and Y.~Urakawa, \emph{{Relaxion window}},
  \href{https://arxiv.org/abs/1605.06908}{{\ttfamily 1605.06908}}.

\bibitem{2015JHEP...12..162B}
B.~{Batell}, G.~{Giudice} and M.~{McCullough}, \emph{{Natural heavy
  supersymmetry}}, \href{https://doi.org/10.1007/JHEP12(2015)162}{\emph{Journal
  of High Energy Physics} {\bfseries 12} (Dec., 2015) 162},
  [\href{https://arxiv.org/abs/1509.00834}{{\ttfamily 1509.00834}}].

\bibitem{Weinberg:1987dv}
S.~Weinberg, \emph{{Anthropic Bound on the Cosmological Constant}},
  \href{https://doi.org/10.1103/PhysRevLett.59.2607}{\emph{Phys. Rev. Lett.}
  {\bfseries 59} (1987) 2607}.

\bibitem{Gupta2016}
R.~S. Gupta, Z.~Komargodski, G.~Perez and L.~Ubaldi, \emph{Is the relaxion an
  axion?}, \href{https://doi.org/10.1007/JHEP02(2016)166}{\emph{Journal of High
  Energy Physics} {\bfseries 2016} (Feb, 2016) 166}.

\bibitem{Choi:2015fiu}
K.~Choi and S.~H. Im, \emph{{Realizing the relaxion from multiple axions and
  its UV completion with high scale supersymmetry}},
  \href{https://doi.org/10.1007/JHEP01(2016)149}{\emph{JHEP} {\bfseries 01}
  (2016) 149}, [\href{https://arxiv.org/abs/1511.00132}{{\ttfamily
  1511.00132}}].

\bibitem{Kaplan:2015fuy}
D.~E. Kaplan and R.~Rattazzi, \emph{{Large field excursions and approximate
  discrete symmetries from a clockwork axion}},
  \href{https://doi.org/10.1103/PhysRevD.93.085007}{\emph{Phys. Rev.}
  {\bfseries D93} (2016) 085007},
  [\href{https://arxiv.org/abs/1511.01827}{{\ttfamily 1511.01827}}].

\bibitem{2005pfc..book.....M}
V.~{Mukhanov}, \emph{{Physical Foundations of Cosmology}}.
\newblock Nov., 2005,
  \href{https://doi.org/10.2277/0521563984}{10.2277/0521563984}.

\bibitem{2016AA...594A..13P}
{Planck Collaboration}, P.~A.~R. {Ade}, N.~{Aghanim}, M.~{Arnaud},
  M.~{Ashdown}, J.~{Aumont} et~al., \emph{Planck 2015 results. xiii.
  cosmological parameters},
  \href{https://doi.org/10.1051/0004-6361/201525830}{\emph{Astronomy \&
  Astrophysics} {\bfseries 594} (Sept., 2016) A13},
  [\href{https://arxiv.org/abs/1502.01589}{{\ttfamily 1502.01589}}].

\bibitem{PhysRevD.92.092003}
J.~M. Pendlebury, S.~Afach, N.~J. Ayres, C.~A. Baker, G.~Ban, G.~Bison et~al.,
  \emph{Revised experimental upper limit on the electric dipole moment of the
  neutron}, \href{https://doi.org/10.1103/PhysRevD.92.092003}{\emph{Phys. Rev.
  D} {\bfseries 92} (Nov, 2015) 092003}.

\bibitem{Spradlin:2001pw}
M.~Spradlin, A.~Strominger and A.~Volovich, \emph{{Les Houches lectures on de
  Sitter space}},  in \emph{{Unity from duality: Gravity, gauge theory and
  strings. Proceedings, NATO Advanced Study Institute, Euro Summer School, 76th
  session, Les Houches, France, July 30-August 31, 2001}}, pp.~423--453, 2001,
  \href{https://arxiv.org/abs/hep-th/0110007}{{\ttfamily hep-th/0110007}},
  \href{http://alice.cern.ch/format/showfull?sysnb=2277605}{http://alice.cern.ch/format/showfull?sysnb=2277605}.

\bibitem{Gross:1980br}
D.~J. Gross, R.~D. Pisarski and L.~G. Yaffe, \emph{{QCD and Instantons at
  Finite Temperature}},
  \href{https://doi.org/10.1103/RevModPhys.53.43}{\emph{Rev. Mod. Phys.}
  {\bfseries 53} (1981) 43}.

\bibitem{Espinosa:2007qp}
J.~R. Espinosa, G.~F. Giudice and A.~Riotto, \emph{{Cosmological implications
  of the Higgs mass measurement}},
  \href{https://doi.org/10.1088/1475-7516/2008/05/002}{\emph{JCAP} {\bfseries
  0805} (2008) 002}, [\href{https://arxiv.org/abs/0710.2484}{{\ttfamily
  0710.2484}}].

\bibitem{Moroi:1993mb}
T.~Moroi, H.~Murayama and M.~Yamaguchi, \emph{{Cosmological constraints on the
  light stable gravitino}},
  \href{https://doi.org/10.1016/0370-2693(93)91434-O}{\emph{Phys. Lett.}
  {\bfseries B303} (1993) 289--294}.

\bibitem{Kawasaki:1994af}
M.~Kawasaki and T.~Moroi, \emph{{Gravitino production in the inflationary
  universe and the effects on big bang nucleosynthesis}},
  \href{https://doi.org/10.1143/PTP.93.879}{\emph{Prog. Theor. Phys.}
  {\bfseries 93} (1995) 879--900},
  [\href{https://arxiv.org/abs/hep-ph/9403364}{{\ttfamily hep-ph/9403364}}].

\bibitem{Bolz:2000fu}
M.~Bolz, A.~Brandenburg and W.~Buchmuller, \emph{{Thermal production of
  gravitinos}}, \href{https://doi.org/10.1016/S0550-3213(01)00132-8,
  10.1016/j.nuclphysb.2007.09.020}{\emph{Nucl. Phys.} {\bfseries B606} (2001)
  518--544}, [\href{https://arxiv.org/abs/hep-ph/0012052}{{\ttfamily
  hep-ph/0012052}}].

\bibitem{German:2001tz}
G.~German, G.~G. Ross and S.~Sarkar, \emph{{Low scale inflation}},
  \href{https://doi.org/10.1016/S0550-3213(01)00258-9}{\emph{Nucl. Phys.}
  {\bfseries B608} (2001) 423--450},
  [\href{https://arxiv.org/abs/hep-ph/0103243}{{\ttfamily hep-ph/0103243}}].

\bibitem{Choi:2016luu}
K.~Choi and S.~H. Im, \emph{{Constraints on Relaxion Windows}},
  \href{https://doi.org/10.1007/JHEP12(2016)093}{\emph{JHEP} {\bfseries 12}
  (2016) 093}, [\href{https://arxiv.org/abs/1610.00680}{{\ttfamily
  1610.00680}}].

\bibitem{Evans:2017bjs}
J.~L. Evans, T.~Gherghetta, N.~Nagata and M.~Peloso, \emph{{Low-scale D -term
  inflation and the relaxion mechanism}},
  \href{https://doi.org/10.1103/PhysRevD.95.115027}{\emph{Phys. Rev.}
  {\bfseries D95} (2017) 115027},
  [\href{https://arxiv.org/abs/1704.03695}{{\ttfamily 1704.03695}}].

\bibitem{LINDE1986395}
A.~Linde, \emph{Eternally existing self-reproducing chaotic inflanationary
  universe},
  \href{https://doi.org/https://doi.org/10.1016/0370-2693(86)90611-8}{\emph{Physics
  Letters B} {\bfseries 175} (1986) 395 -- 400}.

\bibitem{PhysRevD.50.730}
J.~Garc\'{\i}a-Bellido, A.~Linde and D.~Linde, \emph{Fluctuations of the
  gravitational constant in the inflationary brans-dicke cosmology},
  \href{https://doi.org/10.1103/PhysRevD.50.730}{\emph{Phys. Rev. D} {\bfseries
  50} (Jul, 1994) 730--750}.

\bibitem{PhysRevD.52.3365}
A.~Vilenkin, \emph{Making predictions in an eternally inflating universe},
  \href{https://doi.org/10.1103/PhysRevD.52.3365}{\emph{Phys. Rev. D}
  {\bfseries 52} (Sep, 1995) 3365--3374}.

\bibitem{PhysRevD.60.023501}
J.~Garriga, T.~Tanaka and A.~Vilenkin, \emph{Density parameter and the
  anthropic principle},
  \href{https://doi.org/10.1103/PhysRevD.60.023501}{\emph{Phys. Rev. D}
  {\bfseries 60} (May, 1999) 023501}.

\bibitem{PhysRevD.78.063520}
A.~De~Simone, A.~H. Guth, M.~P. Salem and A.~Vilenkin, \emph{Predicting the
  cosmological constant with the scale-factor cutoff measure},
  \href{https://doi.org/10.1103/PhysRevD.78.063520}{\emph{Phys. Rev. D}
  {\bfseries 78} (Sep, 2008) 063520}.

\bibitem{PhysRevD.51.429}
J.~Garc\'{\i}a-Bellido and A.~Linde, \emph{Stationarity of inflation and
  predictions of quantum cosmology},
  \href{https://doi.org/10.1103/PhysRevD.51.429}{\emph{Phys. Rev. D} {\bfseries
  51} (Jan, 1995) 429--443}.

\bibitem{PhysRevD.52.6730}
J.~Garc\'{\i}a-Bellido and A.~Linde, \emph{Stationary solutions in brans-dicke
  stochastic inflationary cosmology},
  \href{https://doi.org/10.1103/PhysRevD.52.6730}{\emph{Phys. Rev. D}
  {\bfseries 52} (Dec, 1995) 6730--6738}.

\end{thebibliography}\endgroup
 
\end{document}